\documentclass[dvips]{article}
\usepackage{icrctc07}
\newcommand{\gr}{$\gamma$-ray \,}

\title{Comparing a model of cosmic ray production in the supernova
remnant\\ 
RX~J1713.7-3946 with observations
}
\shorttitle{Cosmic ray production in RX~J1713.7-3946}
\authors{E.G. Berezhko$^{1}$, H.J. V\"olk$^{2}$.}
\shortauthors{E.G. Berezhko and H.J. V\"olk}
\afiliations{$^1$Yu.G. Shafer Institute of Cosmophysical Research and Aeronomy,
             31 Lenin Ave., 677980 Yakutsk, Russia \\
             $^2$Max-Planck-Institut f\"ur Kernphysik, Postfach 103980, 69029
             Heidelberg, Germany}
\email{berezhko@ikfia.ysn.ru}

\abstract{%

Explicitly time-dependent, nonlinear kinetic theory of cosmic ray (CR)
acceleration in supernova remnants (SNRs) has been employed to investigate the
properties of SNR RX~J1713.7-3946. Observations of the nonthermal radio and
X-ray emission spectra as well as earlier H.E.S.S. measurements of the very
high energy $\gamma$-ray emission were used to constrain the astronomical and
the particle acceleration parameters of the system. The model assumes that the
object was a core collapse supernova (SN) with a massive progenitor, has an age
of $\approx 1600$~yr and is at a distance of $\approx 1$~kpc. It is shown that
an efficient production of nuclear CRs, leading to strong shock modification
and a large downstream magnetic field strength $B_{\mathrm{d}}\sim100$~$\mu$G,
can reproduce the observed synchrotron emission from radio to X-ray frequencies
together with the \gr spectral characteristics as observed by the
H.E.S.S. telescopes. Small-scale filamentary structures observed in nonthermal
X-rays provide empirical confirmation for this field amplification scenario
which leads to a strong depression of the inverse Compton and Bremsstrahlung
fluxes. The results are compared with the latest H.E.S.S. observations.}

\begin{document}
\maketitle
\section{Introduction}

RX~J1713-3946 is an extended shell-type SNR in the Galactic plane that was
discovered in X-rays with ROSAT \cite{pfef96}. It turns out that the
observable X-ray emission is entirely nonthermal. The radio synchrotron
emission is weak, with a poorly known spectral form. RX~J1713-3946 was also
detected in very high energy (VHE) $\gamma$-rays with the CANGAROO
\cite{Muraishi,Enomoto} and H.E.S.S. \cite{Aha04,Aha06}
telescopes. Especially the latter observations show a clear shell structure at
TeV energies which correlates well with the X-ray contours from ASCA. The
different observations are discussed in the paper of Berezhko \& V\"olk
(\cite{BV06}, hereafter referred to as BV06), on which we expand here.

The basic aim of BV06 is an investigation of the acceleration of
both electrons and protons in this remnant, the calculation of the nonthermal
radiation spectrum of the source using an explicitly time-dependent nonlinear
kinetic theory, and a theoretical determination of the (spatial) morphology. In
such a theory the particle acceleration process is coupled with the
hydrodynamics of the thermal gas in the aftermath of the SN explosion.

In comparison with other SNRs (SN 1006, Cas~A, and Tycho's SNR) that were
successfully described within the framework of this theory and finally allowed
a prediction of the \gr flux (see \cite{ber05} for a review), the case of
RX~J1713-3946 is more difficult despite its spatially resolved $\gamma$-ray
emission. First of all, astronomical parameters such as source distance,
present expansion velocity and age are not well known. Still, following
estimates from ASCA observations \cite{Koyama} and from NANTEN CO line
measurements \cite{Fukui,Moriguchi} a rather close distance of $d=1$~kpc
appears indicated, and historical Chinese records suggest a low age of
$t=1614$~yr \cite{Wang}. These parameters are consistent with the assumption of
a core collapse SN of type II/Ib which exploded into the adiabatic, very
diluted bubble created by the wind of a massive ($M<20~M_{\odot}$) progenitor
star, whose likely compact remnant in the center is an identified neutron star
(e.g. \cite{Cassam}).
Secondly, the lack of knowledge of the spectral shape of the radio emission
makes it difficult to derive -- from synchrotron spectral observations -- two
determining physical quantities: the effective strength of the magnetic field
and the proton injection rate into the diffusive acceleration process. Strictly
speaking, it is therefore not possible to theoretically predict the TeV
$\gamma$-ray emission. BV06 argue nevertheless that the observed
overall synchrotron spectral shape, from radio frequencies to the X-ray cutoff,
and the small-scale filamentary structures in the nonthermal X-ray emission of
RX~J1713-3946 are consistent with efficient CR acceleration associated with a
considerably amplified magnetic field. These properties allow in addition a
consistent fit of the observed TeV $\gamma$-ray spectrum. It is strongly
dominated by $\pi^0$-decay emission \cite{Aha06}.
%................................... Fig. 1 ................
\begin{figure*}
  \begin{center}
    \includegraphics [width=\textwidth]{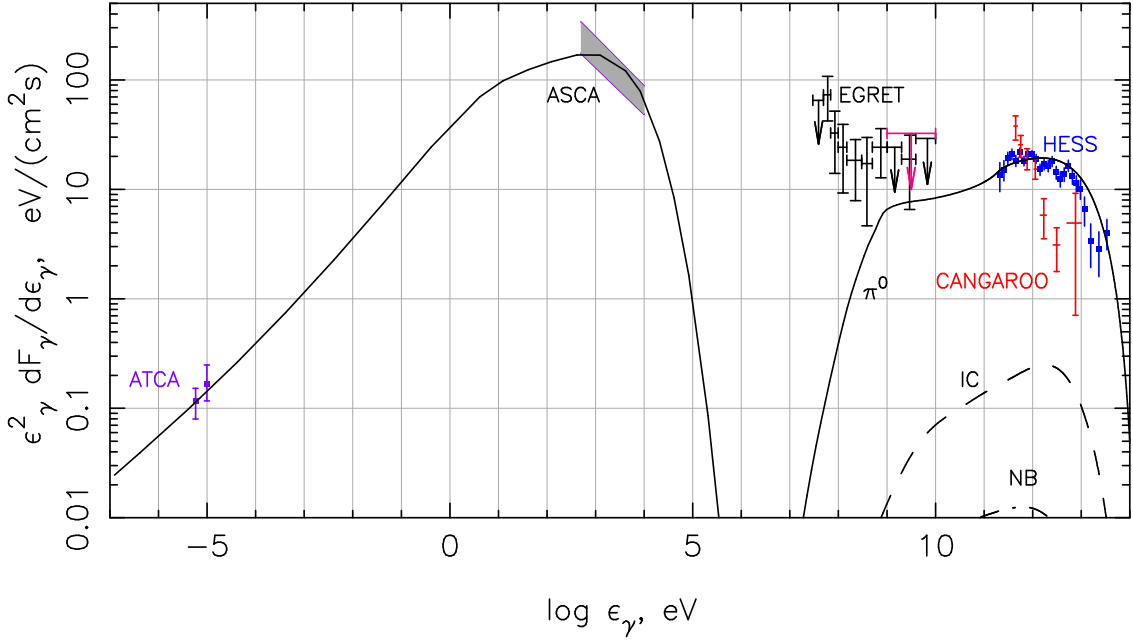} %\hfil
  \end{center}
  \caption{Spatially integrated spectral energy
  distributions of RX~J1713-3946 \cite{BV06}. The {ATCA} radio data, {ASCA}
  X-ray data, {EGRET} spectrum of 3EG~J1714-3857, {CANGAROO} data and 
  {H.E.S.S.} data from (\cite{Aha06}), are shown, together with the {EGRET} 
  upper
  limit for the 
  H.E.S.S. source position.  The solid curve at energies above $10^7$~eV
  corresponds to $\pi^0$-decay \gr emission, whereas the dashed and dash-dotted
  curves indicate the inverse Compton (IC) and Nonthermal Bremsstrahlung (NB)
  emissions, respectively.}
\label{fig1}
\end{figure*}
%............................................................

%................................... Fig. 2 ................
\begin{figure*}
  \begin{center}
    \includegraphics [width=\textwidth]{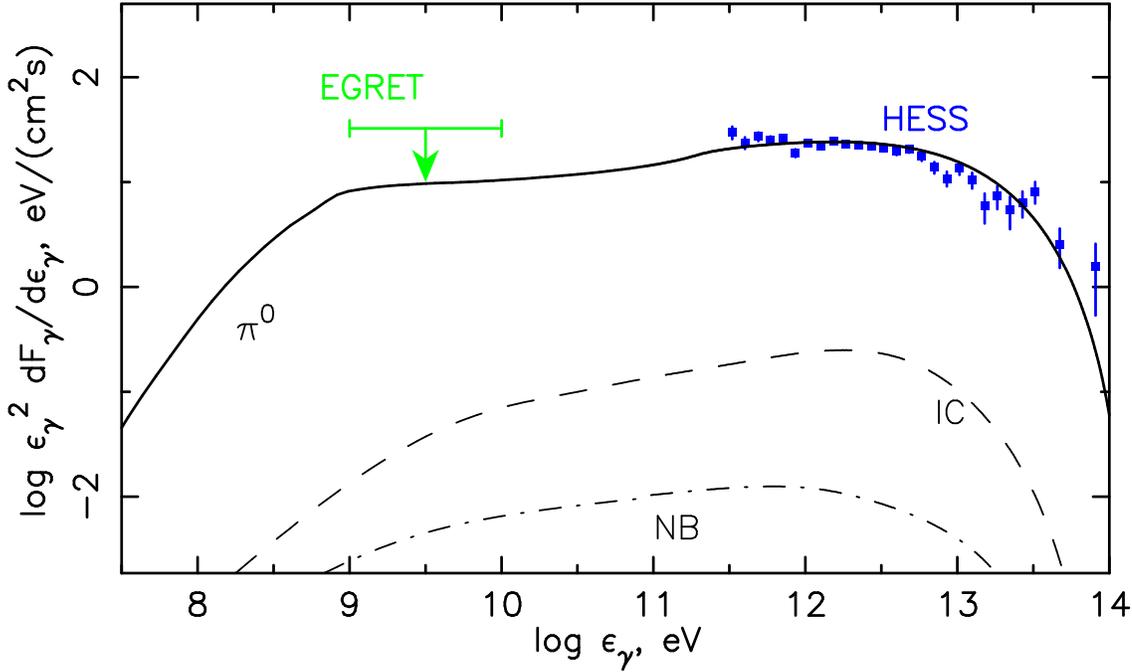} %\hfil
  \end{center}
  \caption{The same theoretical spectra above $10^7$~eV as in Fig.1, 
  on an expanded
  scale, together with the {EGRET} upper limit for the
  H.E.S.S. source position and with the latest H.E.S.S. spectrum from 
  \cite{Aha07}.} 
\label{fig2}
\end{figure*}
%............................................................

\section{Model}

For a present angular radius of the SNR of 60', at the adopted distance of
$d=1$~kpc the linear remnant radius corresponds to 10 pc. To produce the high
observed $\gamma$-ray flux the SN shock must already propagate into the
increasing density $N_{\mathrm H}$ of the shell of swept-up ambient
interstellar medium (ISM) whereas the high nonthermal X-ray emission in the
energy range of several keV requires a presently still high shock speed $V_s
\approx 1840$~km/s, and therefore a very low bubble density $N_b \approx
10^{-2}$~cm$^{-3}$. This requires an ISM density $N_{ISM}=300$~cm$^{-3}$,
i.e. a bubble in a molecular cloud. A consistent hydrodynamic explosion energy
is $E_{\mathrm sn}=1.8 \times 10^{51}$~erg, with an ejected mass of $M_{\mathrm
{ej}} = 3.5~ M_{\odot}$ which give a swept-up mass $M_{sw} \approx 20
M_{\odot}\gg M_{\mathrm {ej}}$, already at the present time.

The effective magnetic field strength $B$ inside the SNR, taken as
approximately uniform \cite{BV04}, determines the strength and form of the
electron distribution for given observed synchrotron emission strength, and
thus also the inverse Compton emission at very high $\gamma$-ray energies for a
given radiation field like the CMB. Since the spatially integrated radio
synchrotron spectrum is not known in detail, $B$ is determined using the
existence of narrow filaments in hard X-rays, whose thickness we interpret as
the result of synchrotron losses of the highest-energy electrons behind the SNR
shock. Recent XMM observations by Hiraga et al. \cite{Hiraga} show such a
narrow filament from which BV06 derived a lower limit to the interior field
strength of $B=65 \mu$G. It is clearly amplified relative to the field strength
in the ambient shell and indicates effective acceleration of nuclear particles
\cite{Bell-Lucek}. The actual value used in the model for the synchrotron
emission is $\approx 130 \mu$G.
%................................... Fig. 2 ................
%\begin{figure}[b]
%  \begin{center}
%    \includegraphics [width=7.1cm]{icrc0614_fig02.eps} %\hfil
%  \end{center}
%  \caption{The same theoretical spectra above $10^7$~eV as in Fig.1, 
%  on an expanded
%  scale, together with the {EGRET} upper limit for the
%  H.E.S.S. source position and with the latest H.E.S.S. spectrum 
%  \cite{Aha07}.} 
% \label{fig2}
% \end{figure}
%............................................................

\section{Results and discussion}

The amplitude of the observed VHE $\gamma$-ray emission can be fitted with a
proton injection rate $\eta = 3\times 10^{-4}$, quite plausible from
theoretical considerations. This is consistent with the high B-field strength
which implies a predominantly hadronic $\gamma$-ray emission, since the
leptonic channel is strongly suppressed for the given synchrotron emission. The
overall shock compression ratio is $\sigma = 6.3$, i.e. the shock is
significantly modified by the accelerated CR protons. It turns out also that
our spherically symmetric acceleration model overproduces nonthermal particle
energy $E_\mathrm{c}$. The necessary deviation of the SNR from spherical
symmetry requires a renormalization $f_{re}E_\mathrm{c}$ with $f_{re}\approx
0.2$, as for a similar object like Cas~A.

Our resulting overall nonthermal spectrum \cite{BV06} is given in Fig.1. In the
VHE range it shows the data from CANGAROO \cite{Enomoto}, together wih the
H.E.S.S. data \cite{Aha06} available at the time. The same theoretical
$\gamma$-ray spectral energy distribution is compared in Fig. 2 with the latest
H.E.S.S. measurements \cite{Aha07} which have significantly increased
statistics and energy coverage. Despite the fact that the theoretical model
does not contain any particle escape at the highest energies, the spectrum
falls off more quickly with energy than the measurement. We trace this back to
the fact that the model of BV06 has -- in a conservative sense -- used the
present value of $B$ as a constant value for $B$ during SNR evolution, whereas
$B^2/(8\pi)$ presumably scales $\propto N_H V_s^2$ \cite{BV04,VBK05} or even
$\propto N_H V_s^3$ \cite{Bell-Lucek}. These latter scalings would lead to a
somewhat higher proton cutoff energy at the present epoch.

Also the theoretical X-ray and TeV $\gamma$-ray morphology can be studied. As
expected from the filament observations, the projected radial X-ray profile
is very narrow. The corresponding profile at 1 TeV is again very narrow,
since in particular the gas density has a large radial gradient. Smoothed to
the resolution of H.E.S.S., the profile is consistent with the observed
H.E.S.S. profile.

A major result remains the hadronic dominance in the $\gamma$-ray emission
spectrum. We expect that the GLAST instrument will confirm our corresponding
theoretical prediction that the spatially integrated $\gamma$-ray spectral
energy density at 1 GeV is only a factor $\approx 2.5$ lower than at 1 TeV.

\section{Acknowledgements}
EGB acknowledges the hospitality of the Max-Plank-Institut f\"ur Kernphysik
where part of this work was carried out. This work has been supported by the
Russian Foundation for Basic Research (grants 05-02-16412, 06-02-96008,
07-02-0221).

%This is the reference to .bib file (Whitout .bib!)
\bibliography{icrc0614_rev}
%This in the bibtex style, is ok.
%\bibliographystyle{plain}
\bibliographystyle{plain}

\end{document}